\documentclass[aps,prb,twocolumn,superscriptaddress,showpacs,amssymb,floatfix]{revtex4-1}
\usepackage{graphicx,bm,xcolor}

\renewcommand{\deg}{\ensuremath{^\circ}}
\newcommand{\xNi}{\ensuremath{x_{\mathrm{Ni}}}}
\newcommand{\alert}[1]{{\color{purple}#1}}

\begin{document}

\title{Ab initio study of pressure stabilised NiTi allotropes: pressure-induced transformations and hysteresis loops}

\author{David Holec}\email{david.holec@unileoben.ac.at}
\affiliation{Department of Physical Metallurgy and Materials Testing, Montanuniversit\"at Leoben, Franz-Josef-Stra\ss{}e 18, A-8700 Leoben, Austria}
\author{Martin Fri\'ak}\email{m.friak@mpie.de}
\affiliation{Max-Planck-Institut f\"{u}r Eisenforschung GmbH,
Max-Planck-Strasse 1, 40237~D\"{u}sseldorf, Germany}
\author{Anton\'{i}n Dlouh\'{y}}
\affiliation{Institute of Physics of Materials, Academy of Sciences of the Czech Republic, \\ \v{Z}i\v{z}kova 22, CZ-616~62~Brno, Czech
Republic}
\author{J\"{o}rg Neugebauer}
\affiliation{Max-Planck-Institut f\"{u}r Eisenforschung GmbH,
Max-Planck-Strasse 1, 40237~D\"{u}sseldorf, Germany}

\date{\today}


\begin{abstract}
Changes in stoichiometric NiTi allotropes induced by 
hydrostatic pressure have been studied employing density functional theory. By modelling the pressure-induced transitions in a way that imitates quasi-static pressure changes, we show that the experimentally observed B19$'$ phase is (in its bulk form) unstable with respect to another monoclinic phase, B19$''$. The lower symmetry of the B19$''$ phase leads to unique atomic trajectories of Ti and Ni atoms (that do not share a single crystallographic plane) during the pressure-induced phase transition. This uniqueness of atomic trajectories is considered  a necessary condition for the shape memory ability. The forward and reverse pressure-induced transition B19$'${$\leftrightarrow$}B19$''$ exhibits a hysteresis that is shown to originate from hitherto unexpected complexity of the Born-Oppenheimer energy surface.
 \end{abstract}



\maketitle

\section{Introduction}

\begin{figure}[b]
  \includegraphics[width=0.9\columnwidth]{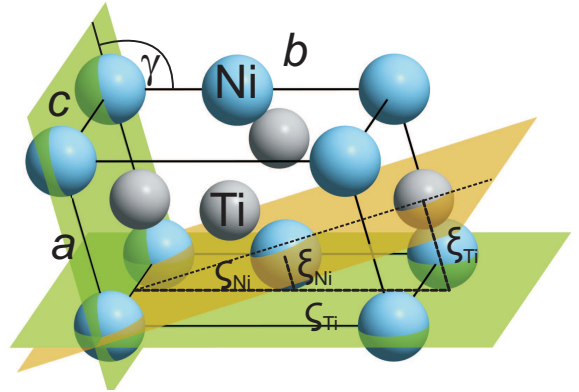}
  \caption{Atomic geometry of the investigated B19$'$-like phases. The various structures considered in this study alternate in the lattice parameters $a$, $b$, $c$, monoclinic angle $\gamma$, and internal positions (see text for details). Larger blue spheres correspond to Ni, smaller gray spheres to Ti atoms. The highlighted planes are used to characterize the structural ability to accommodate the shape-memory effect (see Sec.~\ref{sec:shapeMemory}). The picture was generated using the VESTA package\cite{Momma2008}.}
    \label{fig:B19''}
\end{figure}

Nickel-titanium alloys belong to the important class of shape-memory materials \cite{Hornbogen1991, Saburi1998, Van-Humbeeck1999, Duerig1999}. Their  properties include super-elasticity, excellent mechanical strength and ductility, good corrosion resistance and bio-compatibility (important for example in medical applications), and high specific electric resistance (allowing the material to be easily heated by an electric current). The shape memory effect is governed by a martensitic transformation from a high-temperature austenitic phase (cubic B2, CsCl-structure) into a low-temperature martensitic phase. X-ray experiments on single crystals\cite{Kudoh1985, Michal1981} and neutron measurements on powder samples\cite{Buehrer1983} revealed the low temperature phase to be a monoclinic B19$'$ structure (see Fig.~\ref{fig:B19''}, $\gamma\approx97.8\deg$) with P$2_1$/m space group. In addition, a rhombohedral R-phase\cite{Hara1997} with P3 space group was found during multi-step martensitic transformations\cite{Khalil-Allafi2002, Dlouhy2003, Khalil-Allafi2004, Bojda2005, Michutta2006} under the following conditions: (i)~off-stoichiometric composition, (ii)~presence of substitutional or interstitial impurities, and/or (iii)~formation of precipitate phases.

Several theoretical studies on the low temperature martensitic phase of stoichiometric NiTi alloys have been performed. The intense search has been motivated in part by the fact that theoretically predicted structures do not unambiguously agree with those detected experimentally. For example \citet{Huang2003} concluded that the B19$'$ structure is unstable with respect to a higher-symmetry base-centered orthorhombic (BCO, in some studies also termed B33) structure (see Fig.~\ref{fig:B19''}, $\gamma\approx107\deg$). These conclusions were based on systematically cross-checking several distinct DFT methods, functionals, and implementations (FLAPW, PAW, USPP, GGA, LDA, ABINIT, VASP, etc.). The analysis also considered a carefully selected shear transformation path connecting all three structures B2 ($\gamma=90\deg$), B19$'$, and BCO, since they are characterized by a specific value of the crystallographic angle $\gamma$. Very similar results were reported by \citet{Wagner2008} and by \citet{GudaVishnu2010}. The latter authors\cite{GudaVishnu2010} also predicted a new phase (B19$''$) characterized by $\gamma\approx102.5\deg$ and with practically identical energy to the BCO phase. Finally, a barrier-less transformation path between the B2 and the BCO phases as a sequence of several special deformation modes was demonstrated in Ref.~\onlinecite{GudaVishnu2010}.

Various explanations of the discrepancy between (i)~the apparent stability of the B19$'$ phase as observed in low-temperature experiments and (ii)~the instability of the B19$'$ phase predicted by theoretical calculations (for $T=0\,\mathrm{K}$) have been proposed: Recent theoretical works of \citet{Sestak2011} and \citet{Zhong2011} suggest that the B19$'$ may be stabilized by the presence of (nano)twins that are often experimentally observed\cite{Wagner2008}. As another possibility \citet{Huang2003} suggested that the B19$'$ structure could be stabilized by residual stresses that are frequently present in experimental samples. Since the equilibrium volume is predicted to be smaller for the B19$'$ structure than for the BCO phase\cite{Huang2003}, one may expect the BCO structure to transform into the B19$'$ phase under compressive loads. 

Considering this variety of mechanisms active in NiTi and in order to understand how external strains effect the stability of the various phases, we systematically explore the potential energy surface (PES). To complement previous studies, we focus solely on martensitic phase transformations induced by volumetric changes, i.e., hydrostatic pressure. Our choice is motivated by the fact that (i) stress/strain fields in NiTi alter process-parameters of the martensitic transformations (such as e.g., the transition temperature) and (ii) these actual stresses and strains in experimental samples are difficult to measure and are often not known. Focusing on volumetric changes, we \alert{show} an unexpectedly complex PES. This complexity results in transformation mechanisms that exhibit hysteresis effects not reported in previous studies. From a methodological point of view, we also show that it is difficult to include internal variables explicitly in the PES since they are responsible for metastability of and the newly discovered hysteresis processes.


\section{Computational Details}

The calculations were performed using density functional theory (DFT)\cite{Hohenberg1964,Kohn1965} in the generalized gradient approximation (GGA-PBE'96)\cite{Perdew1996} as implemented in the Vienna Ab-initio Simulation Package (VASP)\cite{Kresse1993,Kresse1996}. All monoclinic structures were studied using  four-atom cells with different external and internal parameters, while a two-atom cell was used for the B2 phase. As the total energy differences among different phases are rather small, it was necessary to ensure convergence of the energy below \alert{$1\,\mathrm{meV}$ per formula unit (f.u.), i.e., one Ni and one Ti atom}. Therefore, the plane wave cutoff energy was set to $400\,\mathrm{eV}$ and a $24\times16\times18$ $\bm{k}$-point Monkhorst-Pack mesh was used to sample the Brillouin zone of the monoclinic allotropes studied.

\subsection{Computational Methodology: Quasi-Static Volumetric Changes}

The  computational approach usually employed for studying the effect of hydrostatic pressure is based on determining the total energy as function of volume. The hydrostatic pressure in the system is obtained by fitting the equation of state\cite{Murnaghan1944} to the calculated energy--volume data points. Because the B19$'$ and the BCO phases are structurally similar and differ only slightly in few internal (atomic coordinates) and external (lattice constants and the angle $\gamma$) parameters, the multi-dimensional Born-Oppenheimer potential energy surface (PES) is expected to be quite complex, exhibiting many local minima. In order to explore the impact of hydrostatic pressure on phase stability and martensitic phase transformations among different NiTi allotropes, we determined the PES as function of (i) the atomic volume, (ii) Ni atom $x$-axis internal coordinate, and monoclinic angle $\gamma$ (see details below). In order to systematically map the complex PES, we adopted a quasi-static (QS) approach, within which the volume is increased/decreased gradually in an adiabatic-like manner (see detailed explanation in Appendix~\ref{app-QS}). This not usually used approach allows for more realistic simulations of gradually increasing/decreasing pressures since it closely imitates experimental conditions.

\section{Results and discussion}

\subsection{The monoclinic allotropes under hydrostatic load}

\begin{figure*}[t]
  \begin{minipage}{\columnwidth}
    {\sffamily\large(a)}\hfill\mbox{}\vspace{-2eX}\par
    \includegraphics[width=0.9\columnwidth]{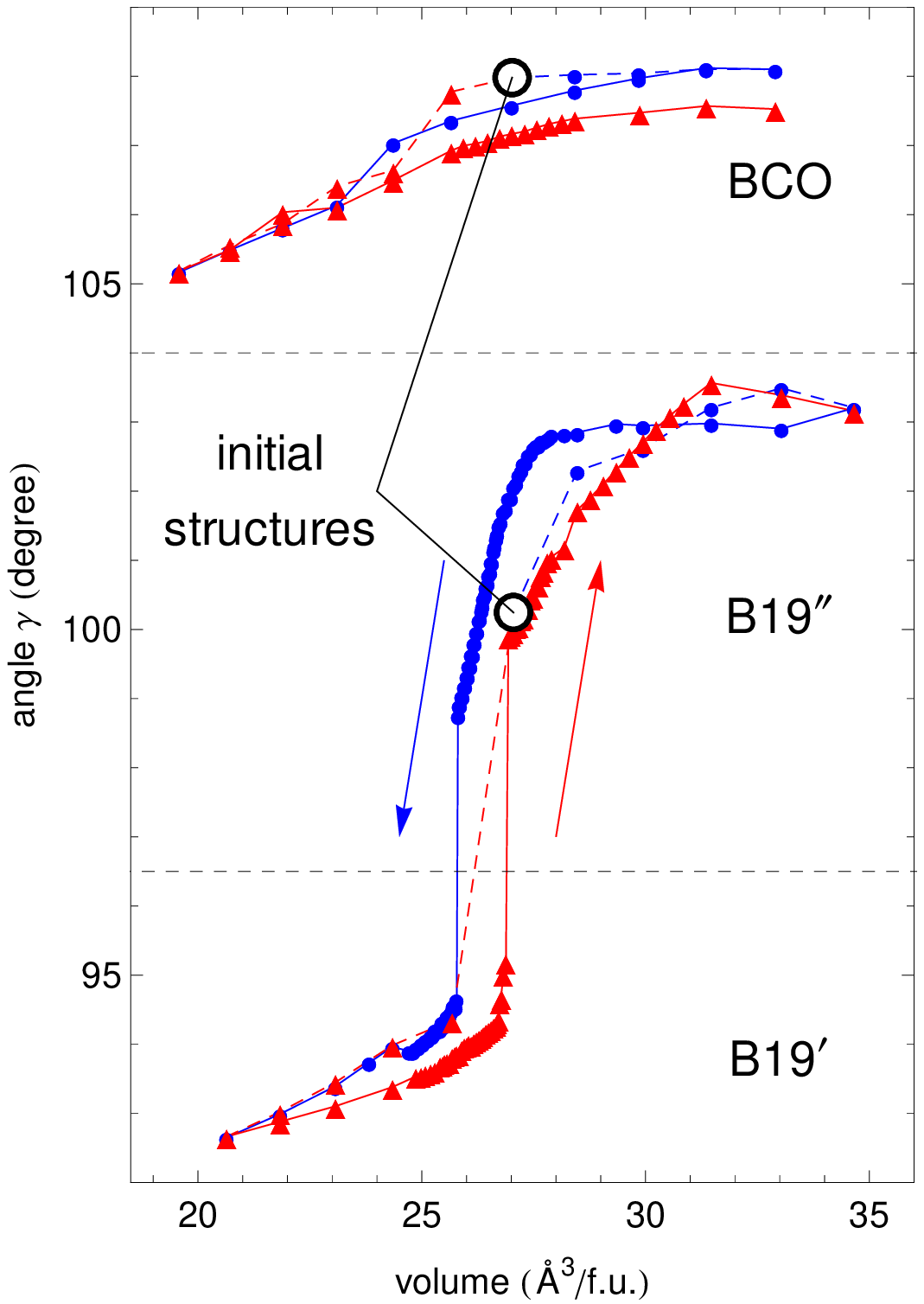}
  \end{minipage}
  \begin{minipage}{\columnwidth}
    {\sffamily\large(b)}\hfill\mbox{}\vspace{-2eX}\par
    \includegraphics[width=0.9\columnwidth]{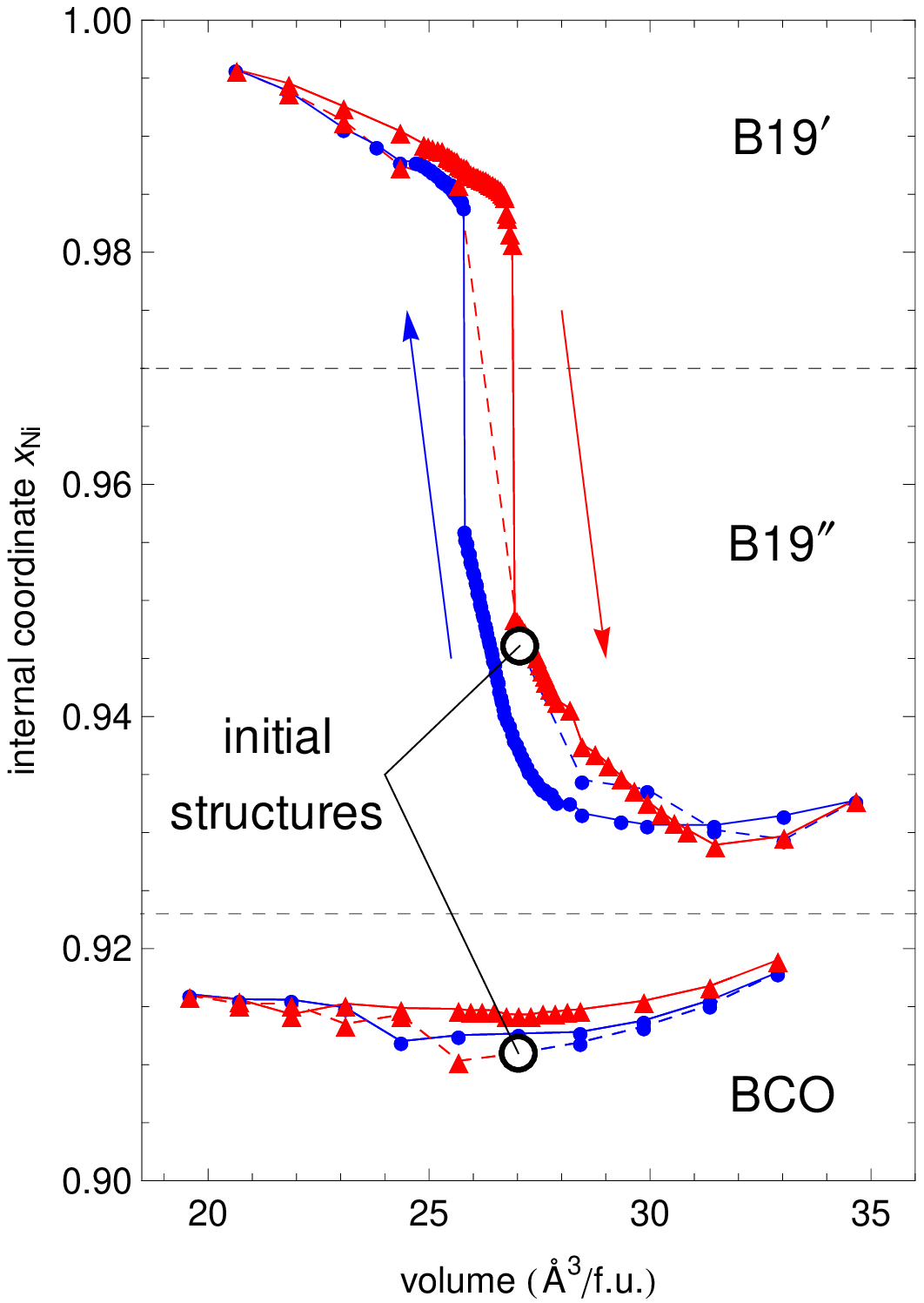}
  \end{minipage}
  \caption{Calculated dependence of (a) the monoclinic angle, $\gamma$, and (b) the internal coordinate, $\xNi$, on the volume obtained from the QS simulations (see  text). Along the B19$'$ compression/decompression paths a region when an additional phase (B19$''$) becomes stable is found. Three ranges of monoclinic angle $\gamma$, that correspond (from the lowest values of $\gamma$ to the highest) to B19$'$, B19$''$ and BCO phases, respectively, are separated by horizontal dashed lines.}
    \label{fig:gamma.and.xNi}
\end{figure*}

The QS simulations were initiated using the previously identified ground states for each phase (B19$'$ and BCO). Subsequently, both structures were to evolve quasi-statically under applied volumetric changes. 
Fig.~\ref{fig:gamma.and.xNi} summarizes results from four separate simulations. From the initial configuration (either  B19$'$ or BCO), we first gradually increased the volume to the maximum value studied here (blue circles and dashed lines in Fig.~\ref{fig:gamma.and.xNi}), and subsequently decreased the volume in the QS manner to the lowest calculated value (blue circles and solid lines in Fig.~\ref{fig:gamma.and.xNi}). Similarly, we proceeded in the opposite direction: from the initial state we first decrease the volume down to the minimum investigated value (red triangles and dashed lines in Fig.~\ref{fig:gamma.and.xNi}), and then increased it to the maximum (red triangles and solid lines in Fig.~\ref{fig:gamma.and.xNi}). We applied these two forward-and-backward runs to both the B19$^\prime$ and BCO starting configurations.

When starting the QS volumetric changes with the BCO phase, the angle $\gamma$ ranges between $105^\circ$ for $20\,\mathrm{\mbox{\AA}^3/f.u.}$ ($\approx60\,\mathrm{GPa}$, compression) and $108^\circ$ for $33\,\mathrm{\mbox{\AA}^3/f.u.}$ ($\approx-20\,\mathrm{GPa}$, expansion). The internal coordinate $\xNi$ remains almost constant at the value $\approx0.915$. In contrast to what was suggested by \citet{Huang2003}, no transition to the B19$'$ phase is observed within this fairly broad range of hydrostatic pressures.

A very different behavior is obtained for the B19$'$ starting configuration. The application of positive hydrostatic pressures (red dashed path in Fig.~\ref{fig:gamma.and.xNi}) first changes the starting angle $\gamma$ abruptly from $\approx100\deg$ to $\approx94\deg$. Further decreasing the volume results in only small changes of the angle $\gamma$. Again, no transition to the BCO structure is predicted. Surprisingly, when negative pressures are applied (volumetric increase, see the dashed blue path in Fig.~\ref{fig:gamma.and.xNi}), the angle $\gamma$ changes to approximately $103\deg$. The resulting unit cell geometry and the internal coordinates no longer correspond to values typical for either the B19$'$ or the BCO state. A similar behavior is demonstrated in Fig.~\ref{fig:gamma.and.xNi}b for the volumetric dependence of the internal coordinate $\xNi$. The structural parameters of this state are very similar to the B19$''$ phase described by \citet{GudaVishnu2010}.

Our results allow to disregard early suggestions of a B19$'${$\leftrightarrow$}BCO transition induced by hydrostatic pressure. Rather, we conclude that hydrostatic pressure, similar to shear deformations\cite{GudaVishnu2010}, transforms B19$'$ into B19$''$. In contrast to monoclinic shear\cite{GudaVishnu2010}, hydrostatic strain does not drive a transition towards BCO. Finally, we find that the BCO phase is stable with respect to the hydrostatic deformations and does not transform to B19$'$ (or B19$''$).

\subsection{Origin of the B19$'${$\leftrightarrow$}B19$''$ hysteresis}

A closer look at the reaction pathways in Fig.~\ref{fig:gamma.and.xNi} reveals the presence of a narrow hysteresis loop. To explain its origin, we have analyzed the PES along both transition paths (resulting from increasing and decreasing volume). We expressed the total energy as \alert{a} function of a single external parameter, volume $V$, and one selected internal parameter, here the $x^\mathrm{Ni}$ position. We chose the latter parameter because, unlike the angle $\gamma$, it can easily be kept constant in available DFT implementation  and it provides clear ranges defining the two phases, B19$'$ and B19$''$ (see Fig.~\ref{fig:gamma.and.xNi} and Table~\ref{tab:ground-state}).

\begin{figure*}[th]
  \begin{minipage}{\columnwidth}
    {\sffamily\large(a)}\hfill\mbox{}\vspace{-2eX}\par
    \includegraphics[width=0.9\columnwidth]{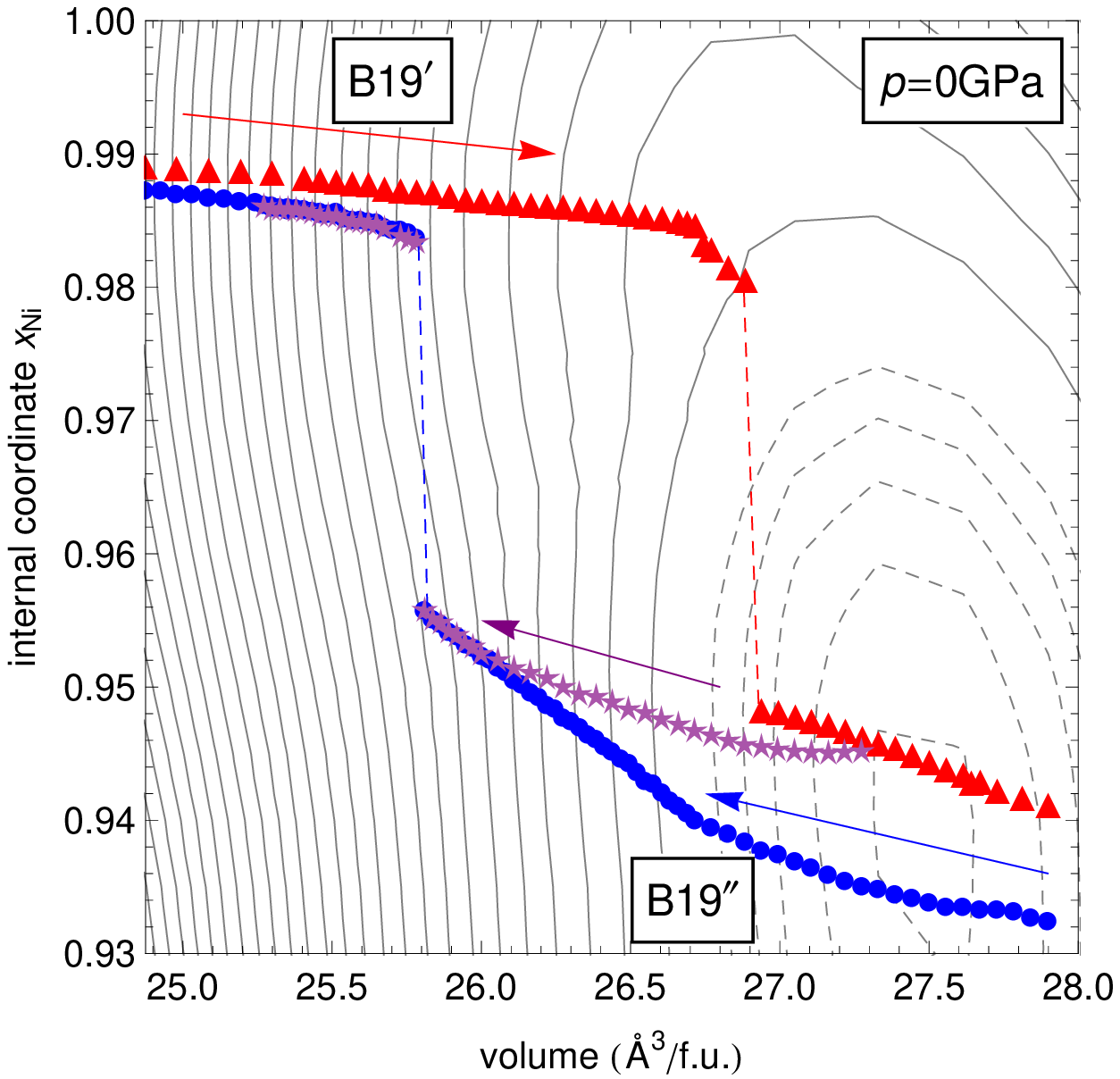}\par

    {\sffamily\large(b)}\hfill\mbox{}\vspace{-2eX}\par
    \includegraphics[width=0.9\columnwidth]{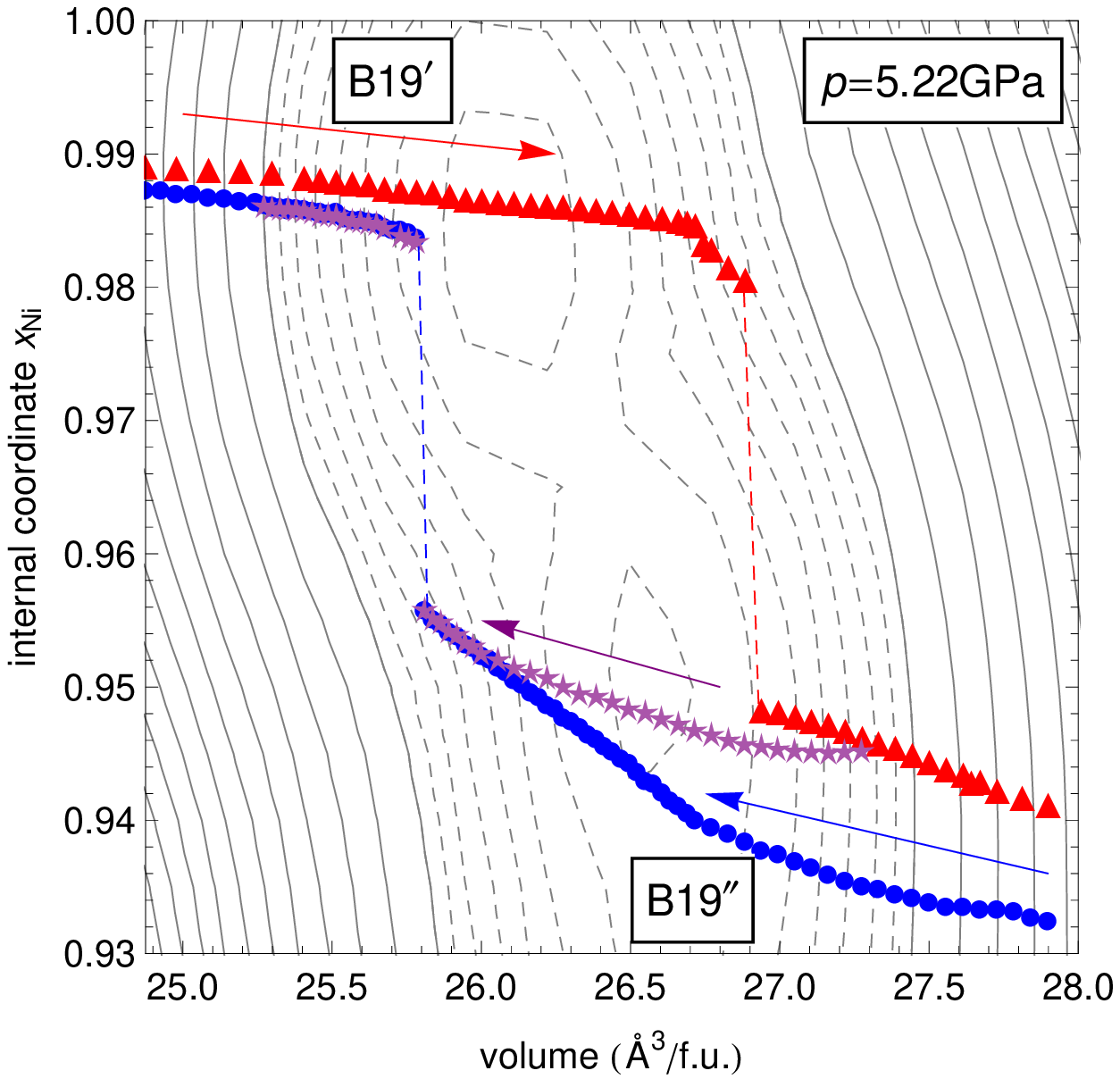}
  \end{minipage}\hfill
  \begin{minipage}{\columnwidth}
    {\sffamily\large(c)}\hfill\mbox{}\vspace{-2eX}\par
     \includegraphics[width=0.9\columnwidth]{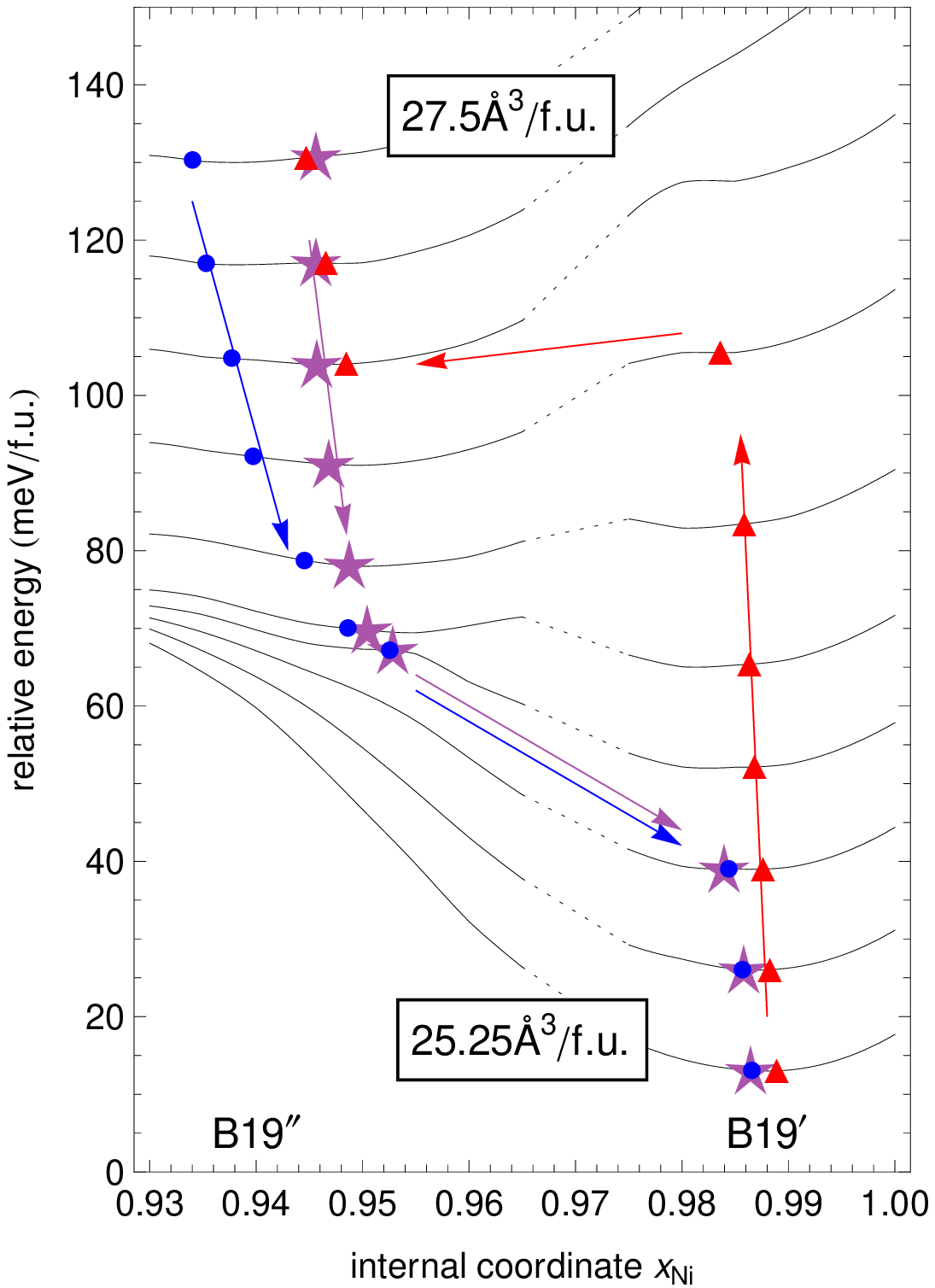}
  \end{minipage}

  \caption{Potential enthalpy surface for $p=0$ for fully relaxed states (a) and for $p=5.22$ GPa (b) along the reaction paths volume $V$ and Ni internal atomic position $x^{\mathrm{Ni}}$. The red, blue and violet symbols mark different volumetric loading conditions and corresponding trajectories (indicated by arrows). Part (c) shows vertical constant-volume cuts of the $p=5.22$ GPa enthalpy.}
    \label{fig:energy_landscape}
\end{figure*}

Using these two parameters we have calculated the total potential energy surface $E^{\mathrm{PES}}(x^{\mathrm{Ni}}, V)$ (Fig.~\ref{fig:energy_landscape}a). As expected, the B19$''$ structure is a stable phase (a global minimum at $V\approx27.5\,\mathrm{\mbox{\AA}^3/f.u.}$ and $\xNi\approx0.94$), while B19$'$ is not associated with any minimum, indicating that in a fully relaxed environment (hydrostatic pressure $p = 0$) this phase is unstable. To investigate the influence of external strain we consider the enthalpy  $H (x^{\mathrm{Ni}}, V, p) = E^{\mathrm{PES}}(x^{\mathrm{Ni}}, V) + pV$  with $p$ being the (hydrostatic) external applied pressure. Increasing the pressure $p$ shifts the equilibrium volume of the  B19$''$ phase towards smaller values (Fig.~\ref{fig:energy_landscape}b). In addition, at sufficiently high pressures, a new minimum occurs that represents for $p=5.22$ GPa the B19$'$ phase. Fig.~\ref{fig:energy_landscape}b explains also neatly the occurrence of the hysteresis. To go from one phase to the other, even at the critical pressure ($p=5.22$ GPa) where both phases have identical enthalpy, a barrier along the constant-volume paths exists. Since in an adiabatic transformation only the nearest local minimum is reachable, the trajectory follows the original path even though this minimum is no longer the energetically most favorable one.

To demonstrate this further,  we plot in Fig.~\ref{fig:energy_landscape}c the energy profiles at fixed volumes (vertical profiles corresponding to the PES in Fig.~\ref{fig:energy_landscape}a). The figure clearly shows that upon increasing the volume (i.e., following the red triangles), the structure is trapped in a local energy valley, and transforms to B19$''$ only when the energy barrier completely flattens. A similar mechanism happens also in the opposite direction (i.e., following the blue circles). 

To further confirm this hypothesis, we performed an additional test. We started from a B19$''$-like structure, but from a volume ($\approx27.3\,\mathrm{\mbox{\AA}^3/f.u.}$) only slightly larger than that at which the B19$'${$\to$}B19$''$ transition occurs ($\approx26.9\,\mathrm{\mbox{\AA}^3/f.u.}$). This pathway is marked by purple stars in Fig.~\ref{fig:energy_landscape}. This pathway also crosses the B19$''$ minimum and eventually joins the B19$''${$\to$}B19$'$ branch of the original hysteresis, i.e., the one corresponding to volume compression (``blue circle'' data points).

From Fig. ~\ref{fig:energy_landscape}b we can further deduce that on changing the transformation coordinates from volume to $x^{\mathrm{Ni}}$ (which is closely related to the monoclinic angle $\gamma$) qualitatively different paths result. In this scenario only a single minimum for a fixed value of $x$ the transformation coordinate (the horizontal cuts of the PES) is obtained, instead of the two-minima for the vertical cuts shown in Fig.~\ref{fig:energy_landscape}c. Consequently, in this case corresponding to a shear mode transformation along the angle $\gamma$, no hysteresis occurs.

We thus conclude that the structural complexity of the B19$'$ and B19$''$ phases and (related to it) the multi-minimum character of the PES are the origin for the transformation hysteresis under hydrostatic loading. This is in contrast to the structurally much more distinct phases, B2 and B19, as shown by \citet{Kibey2009}.

In contrast to what may be expected the volume increasing (red triangles) and decreasing (blue circles) data points in Fig.~\ref{fig:energy_landscape} do not coincide in the region away from the hysteresis loop. The reason is the energy difference between the states (expanding (red triangle) and shrinking (blue circle) volume) at a constant volume is in the order of (or smaller than) $1\,\mathrm{meV/f.u.}$. This value is  below the numerical accuracy of the present calculations. The apparent discrepancy is thus simply a consequence of extremely flat valleys of the PES corresponding to the B19$'$ and, in particular, B19$''$ phases. Increasing the calculation accuracy (albeit at significant CPU costs) is expected to result in a closer correspondence of the two pathways.

\subsection{Structural parameters}\label{sec:structures}

\begin{table*}
  \begin{ruledtabular}
  \begin{tabular}{lcccccccccccc}
    phase & $V_{\rm eq}$ & $B_0$ & $B_0^\prime$ & $\Delta E$ & $a$ & $b$ & $c$ & $\gamma$ & $x^\mathrm{Ni}$ & $y^\mathrm{Ni}$ & $x^\mathrm{Ti}$ & $y^\mathrm{Ti}$\\     
          & $[\mathrm{\mbox{\AA}^3/f.u.}]$ & $[\mathrm{GPa}]$ & & $[\mathrm{meV/f.u.}]$ & $[\mbox{\AA}]$ & $[\mbox{\AA}]$ & $[\mbox{\AA}]$ & & & & & \\ \hline
    B2   & 27.19                 & 160 & 4.00 & 84                  & 3.007                 & 4.253                 & 4.253                 & 90.0$\deg$ & 1.0                   & 0.75                  & 0.5                   & 0.25\\
         & 27.24\footnotemark[1] &     &      & 100\footnotemark[1] & 3.009\footnotemark[1] & 4.255\footnotemark[1] & 4.255\footnotemark[1] & 90.0$\deg${\footnotemark[1]} & 1.0\footnotemark[1]   & 0.75\footnotemark[1]  & 0.5\footnotemark[1]   & 0.25\footnotemark[1]\\
	 &                       &     &      & 92\footnotemark[2]  & 3.014\footnotemark[2] & 4.262\footnotemark[2] & 4.262\footnotemark[2] & 90.0$\deg${\footnotemark[2]} & 1.0\footnotemark[2]   & 0.75\footnotemark[2]  & 0.5\footnotemark[2]   & 0.25\footnotemark[2]\\
    B19$'$ & 26.96                 & 153 & 3.67 & 17                  & 2.732                 & 4.672                 & 4.234                 & 95.3$\deg$ & 0.980                 & 0.823                 & 0.564                 & 0.289 \\
         & 27.52\footnotemark[1] &     &      & 16\footnotemark[1]  & 2.929\footnotemark[1] & 4.686\footnotemark[1] & 4.048\footnotemark[1] & 97.8$\deg${\footnotemark[1]} & 0.953\footnotemark[1] & 0.825\footnotemark[1] & 0.588\footnotemark[1] & 0.283\footnotemark[1]\\
         &                       &     &      & 11\footnotemark[2]  & 2.933\footnotemark[2] & 4.678\footnotemark[2] & 4.067\footnotemark[2] & 98.3$\deg${\footnotemark[2]} & 0.955\footnotemark[2] & 0.826\footnotemark[2] & 0.589\footnotemark[2] & 0.283\footnotemark[2]\\
    BCO  & 27.56                 & 149 & 3.76 & 0                   & 2.914                 & 4.927                 & 4.021                 & 107.3$\deg$ & 0.915                 & 0.829                 & 0.643                 & 0.286\\
         & 27.74\footnotemark[1] &     &      & 0\footnotemark[1]   & 2.940\footnotemark[1] & 4.936\footnotemark[1] & 3.997\footnotemark[1] & 107.0$\deg${\footnotemark[1]} & 0.914\footnotemark[1] & 0.827\footnotemark[1] & 0.642\footnotemark[1] & 0.286\footnotemark[1]\\
         &                       &     &      & 0\footnotemark[2]   & 2.928\footnotemark[2] & 4.923\footnotemark[2] & 4.017\footnotemark[2] & 106.6$\deg${\footnotemark[2]}  & 0.918\footnotemark[2] & 0.829\footnotemark[2] & 0.640\footnotemark[2] & 0.286\footnotemark[2]\\
    B19$''$& 27.43                 & 147 & 5.03 & $<1.0$              & 2.917                 & 4.780                 & 4.047                 & 100.0$\deg$ & 0.945                 & 0.828                 & 0.602                 & 0.284\\
         &                       &     &      & 5\footnotemark[2]   & 2.923\footnotemark[2] & 4.801\footnotemark[2] & 4.042\footnotemark[2] & 102.4$\deg${\footnotemark[2]} & 0.936\footnotemark[2] & 0.829\footnotemark[2] & 0.615\footnotemark[2] & 0.237\footnotemark[2]
  \end{tabular}
  \end{ruledtabular}
  \footnotetext[1]{Ref.~\onlinecite{Huang2003}}
  \footnotetext[1]{Ref.~\onlinecite{GudaVishnu2010}}
  \caption{Calculated external and internal parameters of the B2, B19$'$, BCO, and B19$''$ structures in their ground states. The atomic volumes $V_{\rm eq}$, bulk moduli $B_0$, their pressure derivatives $B_0^\prime$, the total energy differences with respect to the BCO phase of the equilibrium states $\Delta E = E_{\mathrm{eq}}-E^{\rm BCO}_{\rm eq}$, lattice parameters $a$, $b$, $c$, and the angle $\gamma$ together with the internal $x$ and $y$ coordinates of the Ni and Ti atoms predicted in the  present study, and compared with the literature data of \citet{Huang2003} and \citet{GudaVishnu2010}. All the extensive quantities, such as the total energy differences and equilibrium volumes, are listed per two-atom NiTi formula unit (f.u.).}
    \label{tab:ground-state}
\end{table*} 

\begin{figure}[b]
  \centering\includegraphics[width=0.9\columnwidth]{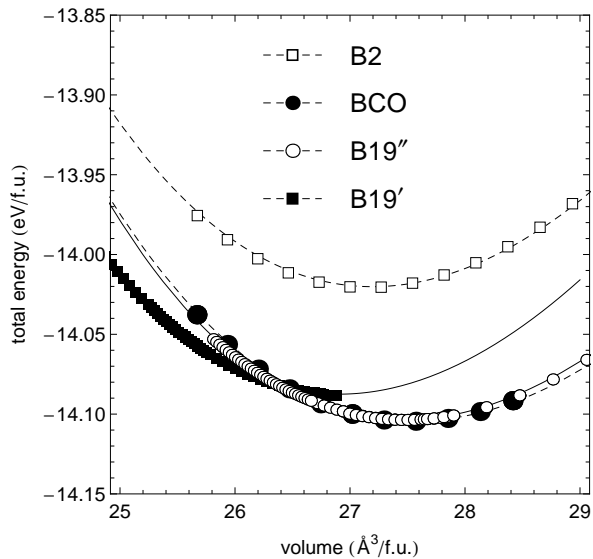}
  \caption{The calculated $E(V)$ curves for the B2, B19$'$, BCO and B19$''$ phases close to the equilibrium.}
    \label{fig:murn_fit}
\end{figure}

The potential energy surface shown in Fig.~\ref{fig:energy_landscape}a provides sets of quasi-static energy--volume data. These data sets that can be individually analyzed using the Murnaghan equation of state\cite{Murnaghan1944}. Following this approach we get the true ground state properties of all phases and can assign a pressure value to each data point. Part of these results are presented in Fig.~\ref{fig:murn_fit}. In this graph all data points are plotted, i.e., from both branches of the hysteresis in Fig.~\ref{fig:gamma.and.xNi}. As a criterion for separating the B19$'$ and B19$''$ phases we used the internal coordinate $x^{\mathrm{Ni}}$: a structure with $x^{\mathrm{Ni}}>0.97$ is B19$'$-like otherwise it corresponds to the B19$''$ phase (see Fig.~\ref{fig:gamma.and.xNi}b). 

Focusing on the most interesting region close to the equilibrium volumes (Fig.~\ref{fig:murn_fit}), we could have easily mistaken the BCO and B19$''$ states as a single phase if we had not performed a thorough analysis of internal coordinates and lattice parameters. The bulk moduli and their pressure derivatives, as well as the equilibrium volumes and the structure energy differences from the Murnaghan equations of state are summarized in Table~\ref{tab:ground-state}, together with all the equilibrium structural parameters. As can be seen, the energy of the B19$''$ state is equal to that of the previously predicted BCO state within the numerical accuracy of our calculations.

A comparison of the structural parameters of our B19$''$ phase with those obtained by \citet{GudaVishnu2010} reveals some differences, the largest being in the monoclinic angle $\gamma$ ($100.0\deg$ predicted here vs. $102.4\deg$ calculated by \citet{GudaVishnu2010}). These are likely to be consequences of using different deformation modes (hydrostatic vs. shear). Despite these small differences we regard these two structures as ``flavors of the same phase and thus use the same name B19$''$ for both of them.

Finally, differences between our structural parameters and those reported in the literature\cite{Huang2003, GudaVishnu2010} of the B19$'$ phase stem from the fact that in the earlier studies the monoclinic angle $\gamma$ was fixed to the experimental value ($\approx98\deg$) while we allowed for a full structural relaxation. As mentioned in the previous section, performing a full relaxation reveals that at ambient pressure the B19$'$ phase is  unstable with respect to the B19$''$ phase.

We note that because some phases are stable only in certain volume (pressure) range, the $E(V)$ data points could not be computed over the whole volumetric range. For example, the B19$'$ phase data points are only available for volumes smaller than $\approx27\,\mathrm{\mbox{\AA}^3/f.u.}$. The predicted properties of the ground states should nevertheless be reasonably accurate as the number of data points obtained is sufficient to perform a numerically robust fitting to the equation of state. 

An advantage of the QS approach is that the energy--volume data points are less scattered (i.e., less influenced by the complexity of the NiTi PES) and their numerical analysis is therefore more robust. Consequently, the initial states (minima of the non-QS energy--volume curves) differ from final states (minima of the quasi-static $E(V)$ curves). An example is shown in Fig.~\ref{fig:gamma.and.xNi} and illustrates these differences: the non-QS value for the monoclinic angle of the B19$'$ structure is $\gamma\approx100.5\deg$ while the QS analysis gives $\gamma\approx94.5\deg$. Since we consider the QS calculations (that mimic the experimental pressure increase or decrease) more accurately, the discrepancy between the QS and non-QS ground states demonstrates the necessity to compute the energy--volume dependence quasi-statically. 

\begin{figure}[t]
  {\sffamily\large(a)}\hfill\mbox{}\vspace{-2eX}\par
  \includegraphics[width=0.9\columnwidth]{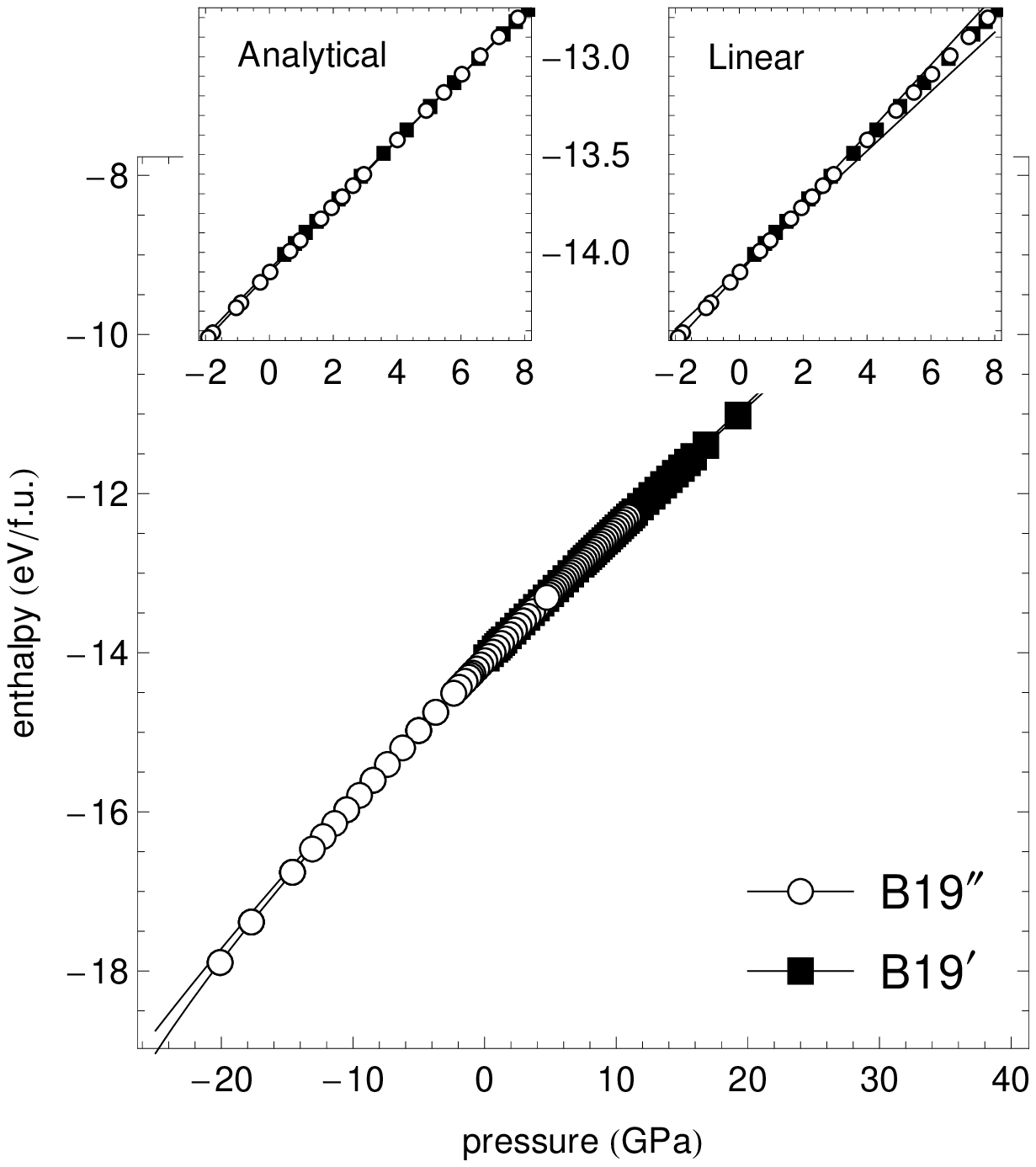}\par

  {\sffamily\large(b)}\hfill\mbox{}\vspace{-2eX}\par
  \includegraphics[width=0.9\columnwidth]{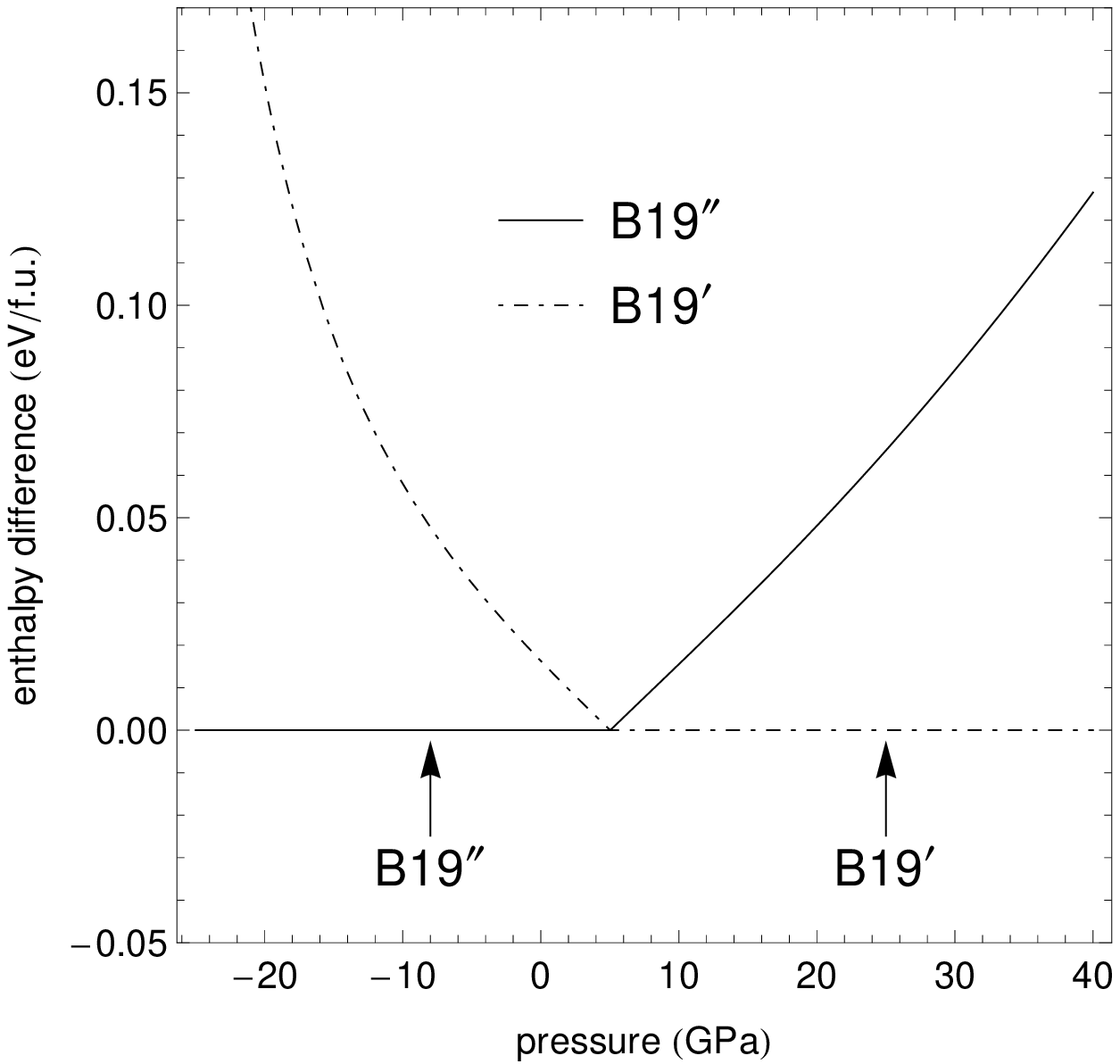}
  \caption{(a) The theoretically predicted enthalpy of the B19$'$ and B19$''$ phases over the whole range of studied pressures. (b) The differences with respect to the phase that minimizes the enthalpy for a given pressure.}
    \label{fig:entalpy}
\end{figure}

In order to determine the critical pressure needed for the B19$'${$\leftrightarrow$}B19$''$ phase transition, we calculated the enthalpies $H$ of both phases  (Fig.~\ref{fig:entalpy}a). Since the enthalpy--pressure data calculated for different phases are similar, an analytical formula for the enthalpy function\cite{Holec2010} was used for the fitting. 

Fig.~\ref{fig:entalpy}b shows the enthalpy difference between a given phase and the phase with the lower enthalpy for a specific pressure. The corresponding transition pressure from B19$''$ to  B19$'$ is $5.22\,\mathrm{GPa}$. 
We note that this pressure dramatically reduces when applying the analytical formula by a linear ($0.47\,\mathrm{GPa}$) or quadratic ($2.01\,\mathrm{GPa}$) fit.  
This  finding clearly demonstrates the necessity of using the analytical expression from Ref.~\onlinecite{Holec2010} based on the Murnaghan equation of state\cite{Murnaghan1944}. It should be further noted that this single-value transition pressure neglects kinetic effects. 

The value of the critical hydrostatic pressure, $\approx5\,\mathrm{GPa}$, above which B19$'$ becomes stable may be compared with the value of $1\, \mathrm{GPa}$ when applying shear stress\cite{Wagner2008}. Exploring the complexity of possible mechanisms active in NiTi alloys, our study and that by \citet{Wagner2008} also complement recent work by \citet{Sestak2011} proposing a twinning mechanism for the stabilization.

\subsection{Ability of the monoclinic allotropes to show a shape memory effect}\label{sec:shapeMemory}

In contrast to the orthorhombic BCO phase (which has a too high symmetry to account for the shape memory effect\cite{Huang2003}), both the B19$'$ and B19$''$ structures possess only a lower (monoclinic) symmetry. The lower symmetry guarantees that the atomic austenite--martensite transition pathway within the unit cell is unique. Thus, the structural phase can, in principle, store the shape information since all the atoms remain situated in the Ericksen-Pitteri neighborhood of their austenite counterparts\cite{Bhattacharya2004}.

In order to recover the BCO lattice from the B19$''$ structure, the Ni atom above the base center has to move into the plane defined by two Ni atoms from the unit cell basal plane and one of the neighboring Ti atoms  (see Fig.~\ref{fig:B19''}). The internal atomic positions then fulfill a specific geometric relation that can be quantified by a structural parameter $\delta$. Employing the internal structural parameters  $\xi_{\rm Ti}$, $\xi_{\rm Ni}$, $\zeta_{\rm Ti}$, and $\zeta_{\rm Ni}$ as defined in Fig.~\ref{fig:B19''}, the parameter $\delta$  is given as:
\begin{equation}
  \delta = \frac{\xi_{\rm Ti}}{\zeta_{\rm Ti}} {\Big /} \frac{\xi_{\rm Ni}}{\zeta_{\rm Ni}}\ .
\end{equation}
When $\delta = 1$, the corresponding Ni and Ti atoms are located within a single plane and the structure is too symmetric to keep an atom-to-atom relationship with the B2 lattice necessary for the shape-memory effect. As the $\delta$ parameter distinguishes whether these Ni and Ti atoms are, or are not, in a planar arrangement, it will be termed ``planarity'' parameter: A deviation from $\delta = 1$ is a prerequisite to store the shape information.

\begin{figure}[t]
  \includegraphics[width=0.9\columnwidth]{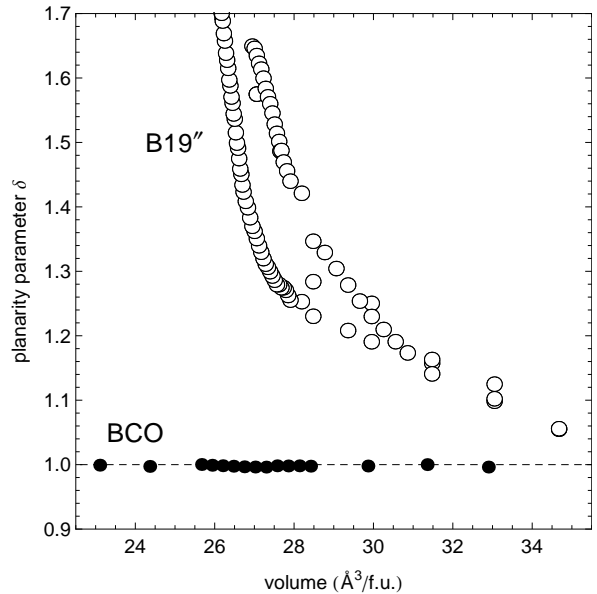}
  \caption{The planarity parameter, $\delta$, describing the internal symmetry of the phases as a function of volume. The value of 1 found for the BCO phase indicates that the Ti atoms are located within the same atomic plane as the Ni atoms (see Fig.~\ref{fig:B19''}) and the internal geometry of the phase is too symmetrical. The B19$''$ values deviating from $\delta=1$ indicate an ability to store the shape information.}
    \label{fig:planeness}
\end{figure}

The planarity parameter $\delta$ for both the BCO and B19$''$ phases is shown in Fig.~\ref{fig:planeness}. Due to the high symmetry environment, the Ti atoms in the BCO phase (full circles in Fig.~\ref{fig:planeness}) are very stable in their location in spite of high hydrostatic pressures (small volumes). $\delta$ is a constant function of volume for the BCO phase indicating that the high symmetry is preserved during the hydrostatic loading. In contrast, the volume dependence of the planarity parameter of B19$''$ (empty circles in Fig.~\ref{fig:planeness}) deviates from  $\delta=1$ over the whole  range of volumes studied. This is in agreement with the analysis of the B19$''$ structure by \citet{GudaVishnu2010}. Within the theory of symmetry-dictated extrema \cite{Craievich1994, Sob1997, Einarsdotter1997, Friak2001, Cerny2005, Friak2008}, the BCO phase represents a structure with a symmetry-dictated energy minimum.

\section{Conclusions}

We report on first-principles calculations of pressure-induced transitions in stoichiometric NiTi allotropes. Complementing previous studies that focused on shear strains and twinning mechanisms, we have systematically explored the complex potential energy surface of NiTi under well-defined generic volumetric changes. We kept the volume constant at each simulation step of the pressure-induced transitions and relaxed all other structural degrees of freedom with respect to the total energy. By repeating these steps in a quasi-static manner, we closely mimicked experimental conditions.

In contrast to previous theoretical studies of shear deformations\cite{Huang2003, Wagner2008, GudaVishnu2010}, the BCO phase does not transform into the experimentally observed B19$'$ phase when applying hydrostatic pressures. We ascribe this stability of the BCO allotrope to the high symmetry of this structure. In contrast, the B19$'$ structure distorts under pressure into another, newly identified, monoclinic phase, B19$''$. This phase is located structurally in between the B19$'$ and BCO phases. We find that the B19$''$ phase has an energy comparable to that of the BCO phase. 

The complexity of the Born-Oppenheimer potential energy surface results in a  pressure-induced B19$'${$\leftrightarrow$}B19$''$ transition that exhibits a previously unreported hysteresis. The latter could be related to the inherent multi-dimensional nature of the potential energy surface in NiTi. The B19$''$ structure   has a lower symmetry than the BCO phase. As a consequence, the B19$''$ structure can be the basis of the shape memory effect.

\section{Acknowledgments}
We thank Dr. Chris Race from the Computational Materials Design department at the Max-Planck Institut f\"{u}r Eisenforschung GmbH in 
D\"{u}sseldorf for carefully reading the manuscript and providing us with helpful comments.

\appendix

\section{The quasi-static approach}\label{app-QS}

In the following we explain the essential difference between quasi-static (QS) and non-QS calculations. The more common non-QS approach can be regarded as a series of energy--volume data points obtained by (i)~starting with an identical internal atomic coordinates and the overall cell shape and (ii)~changing only the overall volume of the unit cell. A so-called relaxed state, i.e., a set of external, $\{\xi_i^n\}$, and internal parameters, $\{\zeta_i^n\}$, for a given volume, $V_n$, is then found by minimizing the total energy, $E$, as a function of both internal and external parameters (except for the volume):
  $$\mbox{non-QS:}\quad\min_{\xi_i=\xi_i^0, \zeta_i=\zeta_i^0} E(V=V_n,\xi_i, \zeta_i) \rightarrow \{\xi_i^n,\zeta_i^n\}$$
where $\xi_i=\xi_i^0, \zeta_i=\zeta_i^0$ reflects the fact that the starting configuration for all $E(V)$ data points (labeled with $n$) is the same. The relaxed states found in the non-QS approach can exhibit phases different from the starting one if pressure-induced structural transitions occur in the studied system. The  non-QS computational approach may fail to properly simulate experimental conditions as the non-QS states that were obtained by discontinuously changed volume may differ from those found in experiments (in which volume and/or strain are always varied continuously). The advantage of the non-QS simulation is that all the calculations can be performed independently in a parallel manner, i.e., calculations for different volumes can be distributed over all available computational units (processors).

In contrast to the non-quasi-static simulations, the quasi-static (QS) simulations can not be parallelized as they proceed by a subsequent series of calculations consisting of the following steps. First, all parameters are energy-relaxed for a certain starting state that frequently corresponds to the equilibrium conditions, i.e., zero hydrostatic pressure. Then, a small change of the volume is applied and new sets of internal and external parameters are obtained by the total energy minimization with respect to the structural parameters (volume being fixed). With these new relaxed parameters (i)~a small volumetric change and (ii)~subsequent structural optimization are repeated. The two steps are then repeated so as to cover the whole range of volumes:
  $$\mbox{QS:}\quad\min_{\xi_i=\xi_i^{n-1}, \zeta_i=\zeta_i^{n-1}} E(V=V_n,\xi_i, \zeta_i) \rightarrow \{\xi_i^n,\zeta_i^n\}\ .$$

The QS simulation mimics a compression of the structure in case of negative volumetric changes and decompression in case of positive ones. The denser is the mesh of the calculated volumes the better correspondence should be achieved with the experimental compression/decompression processes. The QS procedure ensures that the system evolves smoothly from one local minimum into another and the pressure-induced transitions including the phase transition path in a complex configurational space can be studied. In contrast, a non-QS search algorithm may results in (non-physically) discontinuous jumps in the atomic trajectories.


%

\end{document}